# The Influence of Ca and Y on the Microstructure and Corrosion Resistance of Vacuum Die Casting AZ91 Alloy

Feng Wang [1], Jibao Li [1], Jing Liu [1], Pingli Mao [1], Zheng Liu [1]

(1 School of Material Science and Engineering Shenyang University of Technology, Shenyang Liaoning 110870)

**Abstract**: The influence of Ca and Y on the microstructure and corrosion resistance of vacuum die casting AZ91 alloy is investigated using optical microscope, electron scanning microscope, weight-loss test and electrochemical corrosion test. The results indicate that the microstructure of AZ91 alloy can be refined, amount of $Mg_{17}Al_{12}$ phases is reduced, making $Mg_{17}Al_{12}$ phases transform from banding to reticular, and stringer $Al_2Ca$ phases and block $Al_2Y$ phases are formed through adding both Ca and Y. The corrosion resistance of AZ91 magnesium alloy can be increased greatly by adding both Ca and Y. The corrosion rate of AZ91-1.5Ca-1.0Y alloy is dropped to 16.2% of that of AZ91 alloy immersed in 3.5% NaCl aqueous solution for 24 hours. The corrosion current density of AZ91-1.5Ca-1.0Y alloy is dropped by one order of magnitude.

Magnesium alloy is the lightest structural material with high specific strength, specific stiffness, great shock resistance and processing ability, so Mg alloy is widely used in the fields of automobile, aircraft, 3C products and so on[1-4]. But the application of Mg alloy is restricted because of the poor corrosion resistance. Therefore, it is significant to improve the corrosion resistance of Mg alloy. At present, purification treatment, alloying, surface treatment and rapid solidification are mainly used to improve the corrosion resistance of Mg alloy[5,6], and alloying is widely used because of the convenience and obvious results[7,8]. Some researches indicate that the microstructure can be refined, the second phases can be formed and the electrode potential of the matrix can be improved to improve the corrosion resistance through adding Ca, Sr, Ce, Nd and Y elements[9-16]. But little research has been done on the influence of adding both Ca and Y on the corrosion resistance. Therefore, the influence of adding Ca alone and adding both Ca and Y on the microstructure and corrosion resistance of AZ91 alloy is studied in this paper.

## 1 Procedure and Method

Different contents of Ca and Y elements are added into AZ91 alloy to make two different kinds of alloys, and then the tensile test specimens are made by vacuum die casting. The chemical compositions of different alloys are analyzed by Inductively Coupled Plasma (ICP), as shown in table 1.

Table 1 Compositions of alloys (wt. %)

| Alloys | Al | Zn | Ca | Y | Mg |
|---|---|---|---|---|---|
| AZ91 | 9.06 | 0.61 | 0 | 0 | Bal. |
| AZ91-1.5Ca | 8.86 | 0.57 | 1.38 | 0 | Bal. |
| AZ91-1.5Ca-1.0Y | 8.92 | 0.63 | 1.45 | 1.16 | Bal. |

The microstructure of alloy is observed by OLYMPUS BX60 microscope, and the phases and corrosion products are analyzed by Hitachi S-3400 electron scanning microscope (SEM) with an energy-dispersive spectroscopy (EDS). The corrosion rate is tested by weight loss method. The corrosive medium is the solution of 3.5% NaCl and the experiment temperature is room temperature (20℃). Five specimens of each group are weighted (accurate to 0.00001g). The specimens which are soaked in the corrosion solution are swaged with alcohol and acetone to wipe off the corrosion products every 4 hours (the total soaking time is 24 hours), and then washed by water, weighted after dried to calculate the average corrosion rate. The dynamic polarization measurements are tested in CS electrochemical workstation. The platinum electrode is auxiliary electrode, and the saturated electrode is reference electrode. The electrolyte is the solution of 3.5 wt.% NaCl and the experiment temperature is 20℃. The electric potential scanning interval is from Ecorr-200mV to Ecorr+200mV, and the scanning speed is 0.5Mv/s.

## 2 Results and Analysis

### 2.1 The influence of Ca and Y on the microstructure of AZ91 alloy

The metallographic structure, SEM and EDS are shown in figure 1 and 2. Die casting AZ91 alloy is composed of a-Mg and $Mg_{17}Al_{12}$ phase. The a-Mg dendritic crystals are refined and discontinuous strip $Mg_{17}Al_{12}$ phases are distributed among the dendritic crystals (Fig. 1a and Fig. 2a). When 1.5% Ca is added, the microstructure of AZ91 alloy is refined obviously, and the grains are tiny and well-distributed whose average size is about 20μm (Fig. 1b). The EDS analysis result indicates that Ca is partly dissolved in $Mg_{17}Al_{12}$ phases, which make $Mg_{17}Al_{12}$ phases change from strips to sheets and nets. The other Ca is formed into $Al_2Ca$ phases along the grain boundary (shown in figure 2.b, 2.d), and the granular Al-Mn phases are also observed in the microstructure (shown in figure 2.b). When both Ca and Y are added into AZ91 alloy, the microstructure is much more refined and the second phases are much smaller. This indicates that the microstructure of AZ91 alloy can be refined in further when both Ca and Y are added (shown in figure 1.c).

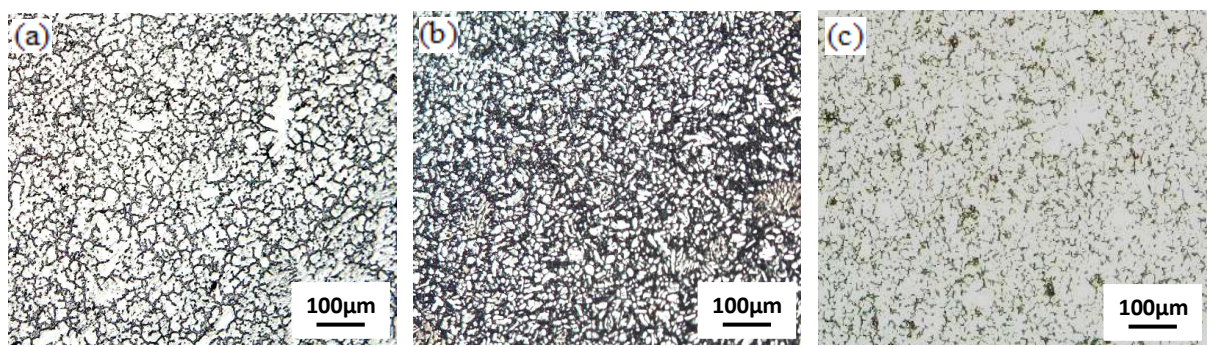

Fig. 1 Microstructures of three alloys: (a) AZ91; (b) AZ91-1.5Ca; (c) AZ91-1.5Ca-1.0Y

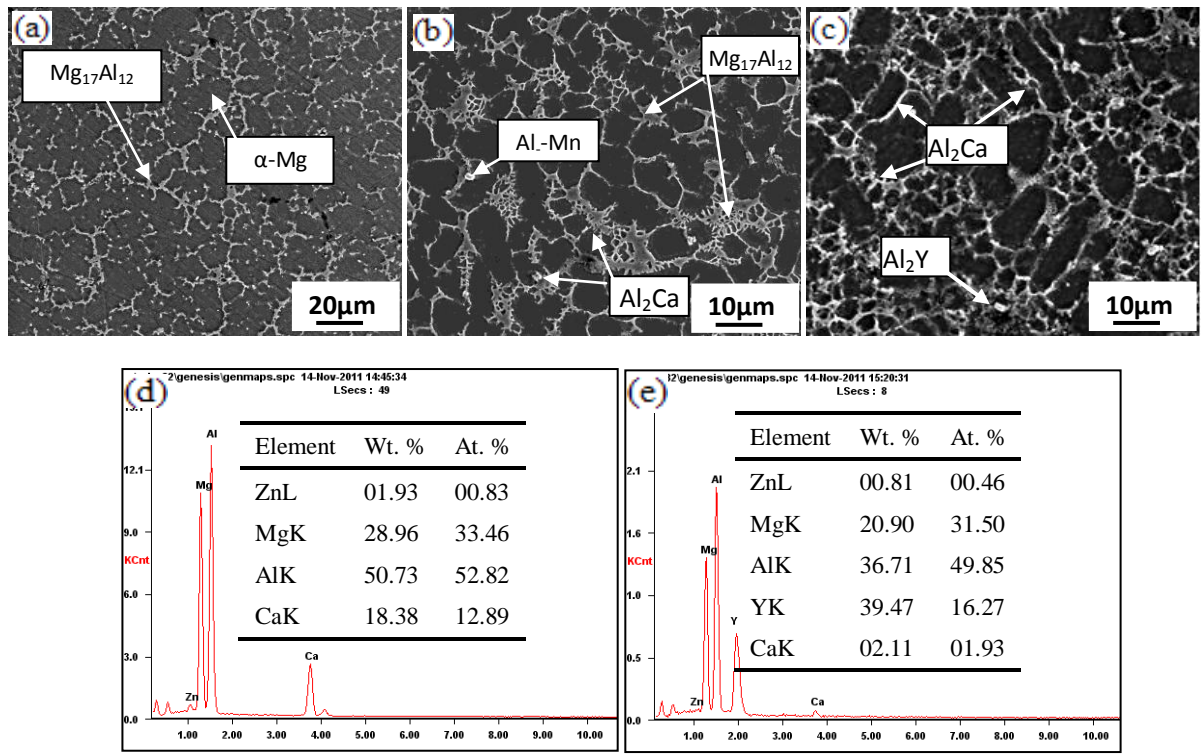

Fig. 2 SEM micrographs of three alloys and EDS patterns of second phase: (a) AZ91; (b) AZ91-1.5Ca; (c) AZ91-1.5Ca-1.0Y; (d) EDS pattern of $Al_2Ca$ phase;(e) EDS pattern of $Al_2Y$

The analysis of XRD of AZ91-1.5Ca-1.0Y alloy is shown in figure 3. As can be seen from figure 3, figure 2.c and figure 2.e, the volume fraction of the second phases is increased obviously after adding 1.0%Y and 1.5%Ca. And the gray strip and net $Mg_{17}Al_{12}$ phases are decreased, while lots of bright white $Al_2Ca$ phases connected with $Mg_{17}Al_{12}$ phases are formed and distributed along the grain boundary in form of net. Some tiny and block $Al_2Y$ phases are formed in the matrix (shown in figure 2.c). This indicates that $Mg_{17}Al_{12}$ phases are decreased, some new second phases are formed and the microstructure is refined and well-distributed.

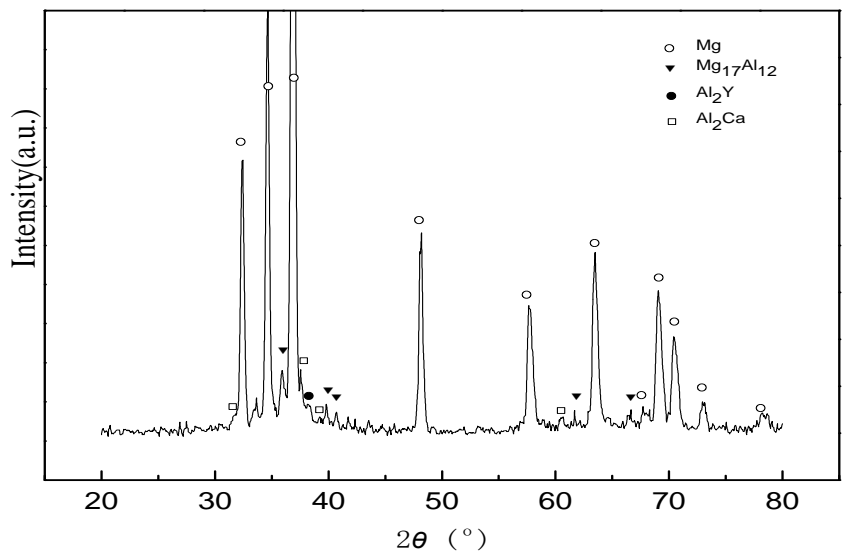

Fig. 3 XRD pattern of AZ91-1.5Ca-1.0Y alloy

## 2.2 The influence of Ca and Y on the corrosion resistance of AZ91 alloy
### 2.2.1 Weight-loss test results and analysis

The curve of influence of the content of Ca and Y and the time of corrosion on AZ91alloy in the solution of 3.5%NaCl tested by weight-loss test is shown in figure 4.

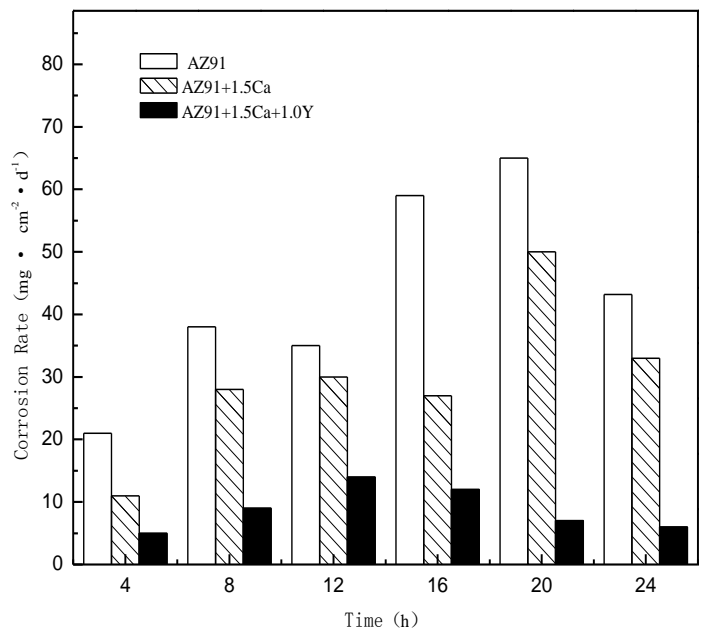

Fig. 4 Effect of Ca、Y content on corrosion rate of AZ91 magnesium alloy

The corrosion rate of AZ91 can be decreased when Ca or both Ca and Y are added. The corrosion resistance of AZ91-1.5Ca-1.0Y alloy is the best in case of the same corrosion time. The corrosion rate of AZ91-1.5Ca-1.0Y alloy which is soaked in the solution of 3.5% NaCl for 24 hours is decreased to 16.2% of that of AZ91 alloy. The corrosion rate of AZ91-1.5Ca-1.0Y alloy reflects an upward trend at the beginning of corrosion but it is decreased after being soaked for 12 hours.

The corrosion rate of the three different kinds of alloys is not always rising or declining but alternative of those. The reason for this is that the corrosion is influenced by two factors. One is that the covering of the surface is formed and thicken, and the other is the pitting corrosion caused by the active anion-$Cl^-$. Because of the poor corrosion resistance of AZ91 alloy, the influence of the active anion is much greater than the cover of the surface. The cover of the surface is destroyed by the pitting corrosion which happens immediately when AZ91 alloy is soaked in the solution. The corrosion tends to be gentle after being soaking for 20 hours, this is probably because of the accumulation of the corrosion products.

### 2.2.2. Polarization Curve

The influence of Ca and Y on the polarization curve of AZ91 alloy soaked in the solution of 3.5%NaCl is shown in figure 5. Corrosion potential can be increased by either adding Ca or adding both Ca and Y, and the corrosion potential of AZ91-1.5Ca-1.0Y is the highest and the corrosion current density is decreased.

The three polarization curves are Tafel fitted and the results are shown in table 2.

The equilibrium potential and the corrosion potential can be increased, and the corrosion current can be decreased by about one order of magnitudes after Ca and Y are added into AZ91 alloy.

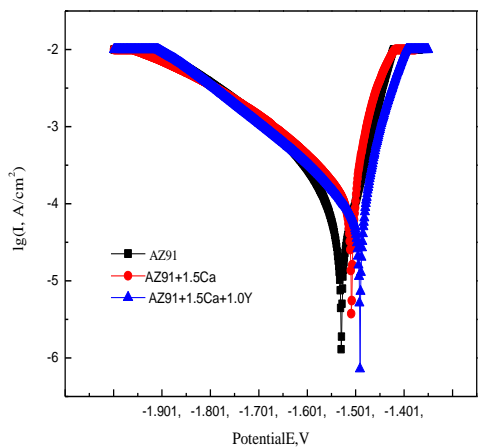

Fig. 5 Polarization curves of three alloys

Table 2 Tafel fitting results for polarization curves of alloys

| Alloys | Equilibrium Potential(V) | Corrosion Potential(V) | Corrosion Current(A) |
|---|---|---|---|
| AZ91 | -1.5214 | -1.5402 | $2.04 \times 10^{-4}$ |
| AZ91-1.5Ca | -1.5043 | -1.5212 | $1.28 \times 10^{-4}$ |
| AZ91-1.5Ca-1.0Y | -1.4975 | -1.4995 | $5.92 \times 10^{-5}$ |

According to the Faraday's law, the corrosion rate is proportional to the current density. Therefore, the corrosion resistance of AZ91-1.5Ca-1.0Y alloy is the greatest among the three alloys. The result is basically in accordance with the change of weight-loss test.

**2.2.3 The Analysis of the Corrosion Products**

In order to recognize the levels of corrosion of alloys with different content of Ca and Y, the alloys are soaked in the solution of 3.5%NaCl for 12 hours, and the macro morphologies of the alloys are shown in figure 6. The surfaces of three alloys are corroded to varying degrees after being soaked for 12 hours. Large corrosion area is shown on the surface of AZ91 alloy, and the corrosion area of AZ91-1.5Ca alloy is smaller. The corrosive degree of AZ91-1.5Ca-1.0Y alloy is the smallest and it is just corroded slightly partly.

The corrosion morphology and EDS of AZ91-1.5Ca-1.0Y alloy soaked in the solution of 3.5%NaCl for 12 hours are shown in figure 7. The main corrosion product is $Mg(OH)_2$, which agree to the reference[8], and the main corrosion place is $Mg_{17}Al_{12}$ phases which are along the grain boundary. The oxide film is damaged by the $Cl^-$.

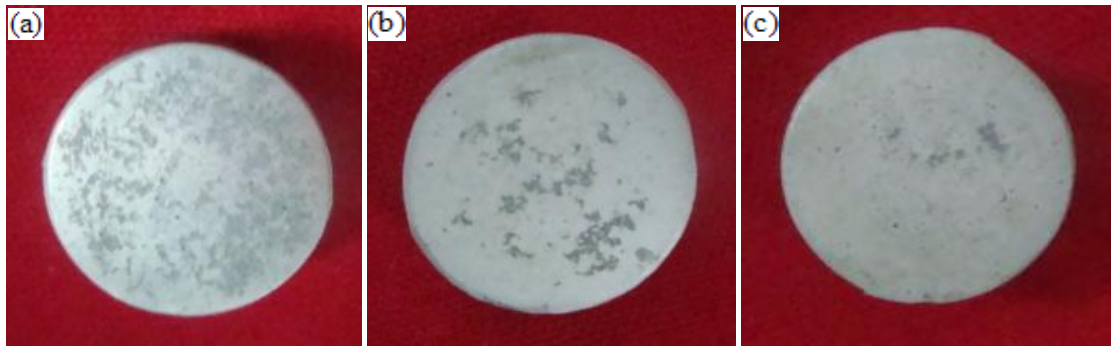

Fig.6 Macro images of three alloys immersed in 3.5% NaCl aqueous solution for 12h:  (a) AZ91; (b) AZ91-1.5Ca; (c) AZ91-1.5Ca-1.0Y

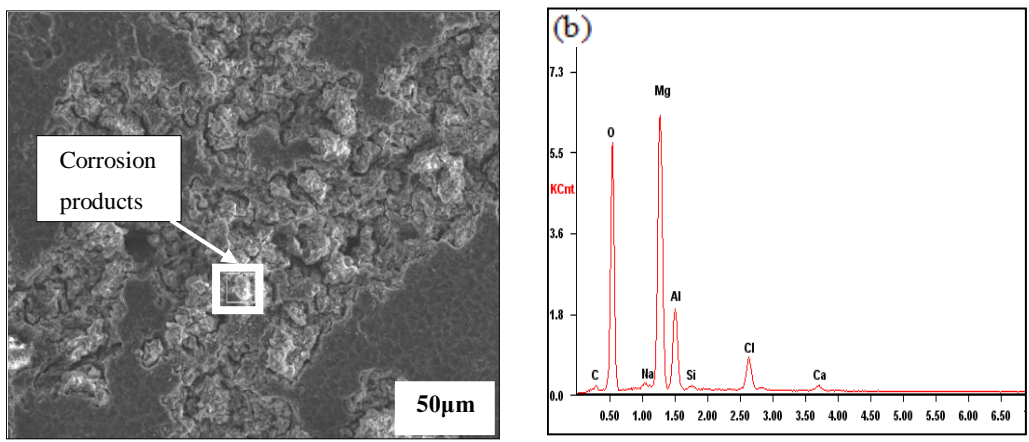

Fig. 7 SEM image (a) EDS pattern (b) AZ91-1.5Ca-1.0Y alloy after corrosion

## 3. Discussion

Ca is a surface-active element, and when Ca is added into AZ91 alloy, Ca is easier to be segregated in front of the solid-liquid interface, so the $Al_2Ca$ phase is alternatively formed. The growth of dendritic crystal is stunted, the microstructure is refined and the distribution of eutectic structure along the grain boundary is more dispersed caused by the $Al_2Ca$ phases. Y is also a surface-active element. The crystallization type is changed because of the increased constituent supercooling caused by Y which is enriched on the solid-liquid interface caused by the limitation of the diffusion dynamics. The microstructure is refined by new phase-$Al_2Y$, which leads to the heterogeneous nucleation and make the growth of β-phase and $Al_2Ca$ stunted.

The corrosion resistance of Mg-Al alloys is influenced greatly by the morphology, size and distribution of a-Mg and $Mg_{17}Al_{12}$. The corrosion resistance of alloy is better when a-Mg is smaller and the amount of $Mg_{17}Al_{12}$ is larger [12-17]. Some researches indicate that part of Al is dissolved in a-Mg and most Al is enriched in $Mg_{17}Al_{12}$ which are along the grain boundary. Al is segregated because of the concentration difference between the grain boundary and inside the grains, and the corrosion resistance of the place with higher content of Al is better. So the corrosion resistance of the grain boundary is better than that of grains [18-19]. The corrosion resistance of the alloys with Ca and Y is increased because the microstructure is refined and the segregation of Al is decreased by Ca and Y.

The content of Al which is dissolved in matrix of AZ91 alloy is 4.58% (wt. %). After Ca and Y are added into AZ91 alloy, the contents of Al in matrix of AZ91-1.5Ca and AZ91-1.5Ca-1.0Y are 1.58% and 3.09% respectively (shown in table 4). This indicates that the corrosion resistance is increased because of the corrosion barrier caused by the distribution of Al along the grain boundary in form of second phase.

The corrosion process of AZ91 alloy is influenced by $Mg_{17}Al_{12}$ in two factors: one is that the corrosion potential of β-phase is higher than that of a-phase, so the corrosion cell is formed by a-phase and β-phase which is the galvanic cathode, and the corrosion resistance is decreased; the other is that a-phase is surrounded by β-phase in form of continuous net and passivation layer is formed by Mg-Al hydroxide and oxide, so the corrosion resistance is increased [20-22]. In this paper, the corrosion resistance of AZ91 alloy is increased by adding Ca or both Ca and Y. $Al_2Ca$ and $Al_2Y$ phases are formed alternatively, so the amount of β-phase which is strong cathode phase is decreased and the morphology of β-phase is changed from strip to net. The corrosion potential is increased and the corrosion current is decreased by the added Ca and Y, so the corrosion rate is decreased. Above all, the corrosion resistance of AZ91 is increased by adding Ca and Y (shown in figure 5). The corrosion resistance of AZ91-1.5Ca-1.0Y is better than that of AZ91-1.5Ca. This is mainly because the growth of $Al_2Ca$ is stunted by $Al_2Y$ which is more stable, the distribution of β-phase is more dispersed and the amount of $Mg1_7Al_{12}$ which is net is decreased. So galvanic corrosion is stunted and the corrosion resistance is increased.

Table 4 EDS analysis of the matrix in different alloys

| 合金/元素 | | Mg | Al | Ca | Y |
|---|---|---|---|---|---|
| AZ91 | wt. % | 95.42 | 4.58 | — | |
| | at. % | 95.85 | 4.15 | — | — |
| AZ91-1.5Ca | wt. % | 98.24 | 1.58 | 0.17 | — |
| | at. % | 98.46 | 1.43 | 0.11 | — |
| AZ91-1.5Ca-1.0Y | wt. % | 96.69 | 3.09 | 0.22 | — |
| | at. % | 97.04 | 2.80 | 0.16 | — |

**4 Results**

(1) The microstructure of AZ91 is refined, $Al_2Ca$ is formed and $Mg_{17}Al_{12}$ is changed from discontinuous strip to sheet and net when 1.5%Ca is added; the microstructure is refined in further and strip $Al_2Ca$ and block $Al_2Y$ are formed when 1.5%Ca and 1.0%Y are added.

(2) The corrosion resistance of AZ91 can be increased by adding Ca and Y. The corrosion rate of AZ91-1.5Ca-1.0Y is the smallest and the corrosion rate of being soaked in the solution of 3.5%NaCl is decreased to the 16.2% of that of AZ91 alloy. The corrosion potential is increased and the corrosion current is decreased by adding both Ca and Y. The corrosion current density is decreased by one order on magnitude.

(3) The microstructure can be refined and distributed more dispersed by adding Ca and Y. The corrosion can be stunted by the net second phases and the corrosion resistance of AZ91-1.5Ca-1.0Y alloy is the greatest.

Acknowledgements
The authors would like to acknowledge the financial support from he "Doctoral Starting up Foundation of Liaoning Province (No:20131083), and the Doctoral Starting up Foundation of Shenyang University of Technology, China.